\documentclass[aps,prb,superscriptaddress,twocolumn,tightenlines,showpacs]
{revtex4}

\usepackage {graphicx}
\usepackage {amsmath}
\usepackage {amsfonts}
\usepackage {amssymb}
\usepackage {mathrsfs}

\newcommand {\rmd} {{\rm d}}

\begin{document}

\title {Topological kinematic constraints: quantum dislocations and the
glide principle.}
\author {V. Cvetkovic}
\email {vladimir@lorentz.leidenuniv.nl}
\affiliation {Instituut Lorentz voor de theoretische natuurkunde, 
Universiteit Leiden, P.O. Box 9506, NL-2300 RA Leiden, 
The Netherlands}
\author {Z. Nussinov}
\affiliation{Department of Physics, Washington University, St. Louis, 
MO 63160-4899, USA}
\affiliation{Los Alamos National Laboratory, MS B256, Los Alamos, NM 87545,
USA}
\email {zohar@viking.lanl.gov}
\author {J. Zaanen}
\email {jan@lorentz.leidenuniv.nl}
\affiliation {Instituut Lorentz voor de theoretische natuurkunde, 
Universiteit Leiden, P.O. Box 9506, NL-2300 RA Leiden, 
The Netherlands}
\affiliation {Department of Physics, Stanford University, CA 94305, USA}

\date {\today}
\begin {abstract}
  Topological defects
play an important role in physics of elastic media 
and liquid crystals. Their kinematics is determined by constraints of 
topological origin. An example is the glide motion of dislocations
which has been extensively studied by metallurgists. In a recent 
theoretical study dealing with quantum dualities associated with
the quantum melting of solids it was argued that these kinematic
constraints play a central role in defining the quantum field theories
of relevance to the description of quantum liquid crystalline states
of the nematic type. This forms the motivation  to analyze more thoroughly
the climb constraints underlying the glide motions. In the setting of
continuum field theory the climb constraint is equivalent to the condition
that the density of constituent particles is vanishing and we derive
a mathematical definition of this constraint which has  a universal
status. This makes possible to study the kinematics of dislocations
in arbitrary space-time dimensions and as an example we analyze the
restricted climb associated with edge dislocations in 3+1D. Very
generally, it can be shown that the climb constraint is equivalent
to the condition that dislocations do not communicate with compressional
stresses at long distances. However, the formalism makes possible 
to address the full non-linear theory of relevance to short distance
behaviors where violations of the constraint become possible.        
\end {abstract}


\maketitle

\section {Introduction}

\begin{figure}
\includegraphics[width=6.5cm]{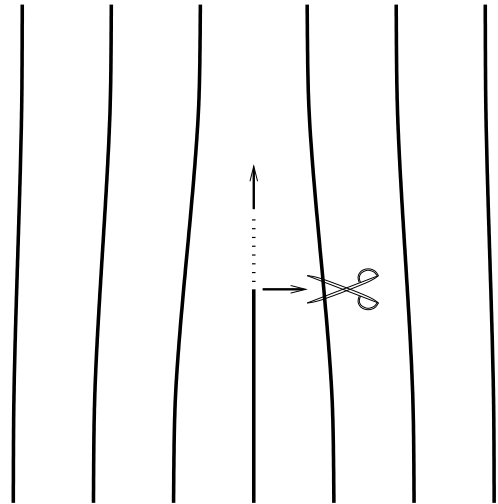}
\caption{An edge dislocation in a two dimensional 
medium. Here, allowed (matter conserving) cutting and reconnecting
of `atomic columns' gives rise to the glide motion of the dislocation line 
(at the center of figure) in the horizontal direction.
By contrast, an emission or absorption of vacancies and interstitials
is required for a climb motion (vertical motion of the dislocation
in this figure). Thus, an absence of additional material (mass conservation) 
strictly prohibits the climb within, continuum limit, 
linear elasticity.} \label {Glidefig}
\end{figure}

Topological defects in crystals 
\cite {topological_defects, kleman, kleman1, mermin,
klemanbook} exhibit a rich variety of kinematic 
properties \cite {Friedel, Nabarro, HirthLothe}. Defects such as
interstitials/vacancies, dislocations, and disclinations 
\cite{Friedel, Nabarro, HirthLothe, Kleinert} 
perturb the ideal perfectly periodic crystal.
To date, numerous works extensively studied 
the dynamics and coarsening 
of these defects, e.g. \cite{Hatano, CB, KHEG, ASV, RR, deem}. In  
normal solids, dislocations are present at low concentrations 
and their peculiar `glide motions' are an important factor in
determining the plastic properties of the medium. It has long been 
recognized that
the energy-entropy balance of topological defects 
is responsible (via deconfinement)
for melting transitions \cite {polyakov, Prls, Onsager, shock, burakovsky, 
KT, Mott, FisherHalperinMorf}. 
In glasses, an extensive configurational
entropy of these defects \cite{glass} 
(from which ensuing restrictive slow 
dynamics might follow \cite{wolynes})  
may be sparked. Investigations of 
Frank Kasper phases \cite{FK1} of defect lines and of
related systems \cite{FK2, FK3, FK4, FK5, FK6, FK7, FK8}
have flourished.
In some solids, such as shape memory alloys 
\cite{shape_memory}, 
reducing the appearance of defects is a matter of pertinence.
The study of topological defects 
in liquid crystals \cite{LC}, compounded by the very crisp imaging techniques
of these systems, 
has evolved into a broad field.
Topological defects often display sharp dynamical imprints. 
Perhaps the best studied such dynamical effect is the glide principle 
which forms the focus of the current article. 
Throughout the years, much work has 
been carried out on the classical glide motion of dislocations. 
Many of these works entail detailed studies of the 
Peach-Kohler forces
between dislocations and Peierls potentials in systems ranging from
classical solids to vortex arrays to liquid crystals, 
e.g. \cite{LS, LBMNR,
RRPBM, MR, SV, AG, NTS}. 
Although the discussion of our current publication is aimed
towards quantum systems, our formalism may be applied
with no change to many of these classical systems to rigorously
derive, in the continuum limit, 
the well known classical glide motion 
(see Fig.(\ref{Glidefig})) occurring in
the absence of interstitials/vacancies and 
high order effects. This rigorous result concerning
the linear regime of continuum limit elasticity complements 
older works addressing detailed climb diffusion 
in various systems, e.g. \cite{bruinsma}. 
We further report on {\em new generalized glide equations}
in the presence of both dislocations and disclinations.
As we illustrate in detail here, the basic physical ingredient leading to
these generalized glide equations in solids is mass conservation
which strictly restrains the dynamics. 

Recent years saw the extension of investi\-gations 
of de\-fect dy\-na\-mics in cla\-ssical media 
to quan\-tum sys\-tems \cite{ZMN} addressing
electronic liquid crystal \cite{KFE} and other phases \cite{ZMN, KFE} 
in which the electronic constituents favor, in a 
certain parameter regime, the formation of an ideal crystal like
stripe pattern which may then be perturbed, through
a cascade of transitions, to produce a rich variety of phases. 
Such stripe patterns are observed in the high temperature
superconductors and other oxides \cite{tranquada}
and in quantum Hall systems \cite{qhe}. 
Defects naturally alter the local electronic density of states allowing
for spatial (and temporal) inhomogeneities 
of electronic properties \cite{zhu}.
Following the general notion that melting 
occurs by the 
condensation of topological defects, 
e.g. \cite{polyakov}, we may naturally anticipate that 
the study of topological 
defects is pertinent to the understanding of 
quantum phase transitions between various zero-temperature states. 
Much unlike classical physics 
where statics and dynamics are decoupled,
in quantum systems, time and space are deeply entangled. As a consequence,
we expect the zero point {\em kinematics} of defects to be of 
paramount importance both to the character of the zero-temperature states 
and to the nature of the phase transitions. 
Only quite recently, along with Mukhin, two of us
\cite {ZMN} presented a thorough
analysis of the problem of topological quantum melting of a solid
in 2+1 space-time dimensions. At first sight, this follows 
the pattern of the famous Nelson-Halperin-Young \cite {NH, Young} 
theory of classical
melting in 2D. However, the quantum problem is far richer. Leaning
heavily on the formalism developed by Kleinert \cite {Kleinert}, 
it was shown in \cite{ZMN} that the transition from elastic solid 
to nematic (or `hexatic')
liquid crystal is closely related to the vortex (or `Abelian Higgs')
duality in 2+1D. The nematic quantum fluid can be viewed as a
Bose-condensate of dislocations subjected to a `dual' Higgs mechanism.
In this formalism, the rigidity of the elastic medium is parameterized
in terms of gauge fields (`stress photons'). In the quantum
fluid, the shear components acquire a Higgs mass due to the presence
of the dislocation superfluid. As it turns out, such a substance is
at the same time a conventional superfluid, which may now be
viewed as an elastic medium which has lost its capacity to sustain
shear stress.

The glide principle as known from metallurgy amounts to the observation
that dislocations only move easily in the direction of their (vectorial)
topological charge. In the nematically ordered state, the
Burgers vectors of the dislocations forming the condensate are
oriented along the macroscopic director leading to an anisotropic
screening of the shear stress \cite {ZMN}. Here, the
dislocations coast freely only
in the direction of their director. Accordingly, the elastic propagator
can only be fully screened in the direction perpendicular to the director. 
Even more striking, it turns out \cite{ZMN} that 
in the quantum field-theoretic 
setting the glide principle acquires a meaning which goes well beyond the
conventional understanding of this phenomenon. In
fact, the main goal of this paper will be to further generalize these
notions beyond the linearized continuum limit of
\cite{ZMN}, including a study of theory in higher 
dimensions and the incorporation of far richer defect configurations.

In many standard texts, 
e.g.\cite {Friedel, Nabarro, HirthLothe}, 
the explanation of glide takes
little effort. A dislocation corresponds with a row of particles 
(atoms) `coming to an end' in the middle of a solid. One way to move 
this entity is to cut the neighboring row at the `altitude' of the
dislocation and consequently move over one 
tail to cure the cut (Fig. \ref {Glidefig}).
The net effect is that the dislocation is displaced. 
This easy mode of motion is termed `glide'. Moving in the 
orthogonal direction is not as easy. Let
us try to move the dislocation `upward'. This requires loose particles
to make the row of particles longer or shorter 
(`interstitials' or `vacancies') and since loose particles are energetically
very costly while they move very slowly (by diffusion) this `climb' motion
is strongly hindered. Estimates on `climb' diffusion rates
are provided in e.g. \cite{bruinsma}.  
Climb is hindered to such an extent in real life
situations (e.g. pieces of steel at room temperature) that it may be ignored
altogether. The lower dimensional glide motion of dislocations
is reminiscent of dynamics in the heavily studied 
sliding phases \cite{slide} in which an effective 
reduction of dimensionality occurs.

Although this simple pictorial explanation suffices for practical matters
in metallurgy, it is immediately obvious that there is more to it.
Dislocations are completely specified by their topological charge, the 
Burgers vector. Expressed in this language, glide is a  
prescription of the dynamics of defects as dictated by 
their own topology: dislocations can only move in the direction of their
Burgers vector! How can this be?

As we review in section II, this kinematic role of topology 
arises most naturally in the context of field theoretic description
of plasticity. Linear elasticity does not capture 
interstitial/vacancy defects. The only degrees of freedom which
are admitted by continuum limit linear elasticity 
are phonons (or, in dualized form, the stress photons).
Given the absence of interstitials, the secret behind 
glide is that
normal crystals are `non-relativistic'-
solids cannot displace along the time direction. As the field
theory lacks knowledge of the interstitials, this condition
translates (as we will illustrate in section
IV) directly into the glide constraint for the dislocation
matter!

From the glide constraint it immediately follows that 
gliding dislocations
do not occupy volume. Volume is exclusively associated with interstitial
matter. Accordingly, dislocations do not communicate with compressional
stress and in the dislocation condensate pressure stays massless because
of the glide constraint! It better be so, because the dislocation condensate
corresponds with a conventional superfluid, and conventional non-relativistic
fluids carry sound. Amusingly, glide needs a preferred time direction and
in a relativistic medium there is no such thing as a preferred direction.
As Kleinert and one of us pointed out \cite {KZ}, 
since glide cannot be defined in a
truly relativistic medium, dislocations have to couple to compressional
stress with the effect that in the relativistic quantum nematic crystal
sound also acquires a mass: this incompressible state turns out to be
nothing else than the space time of 2+1D general relativity.

The main limitation of \cite {ZMN, KZ} 
is that they focus is exclusively on the special
case of dislocation glide in 2+1 dimensions within the realm of the
linearized theory. Our main result, presented in section IV, 
amounts to expressions
for {\em new glide-type constraints in arbitrary dimensions}. 
We further investigate what these constraints imply 
for both dislocations and disclinations via an analogous study
of the more fundamental double curl defect densities. 
In addition, we show in section V
how these wisdoms come to an end in the full non-linear theory
where dislocations can in principle climb, acting as sources and
sinks of interstitials.

This paper is organized as follows:
After reviewing the basics of elasticity and introducing
some notations in section II, we derive expressions for the dynamical
currents for an arbitrarily dimensional medium in section III.
In section IV we deduce
a general relation describing the glide constraint as it acts on
dislocation, disclination, and defect currents according to the linear 
theory, which is applicable in arbitrary dimensions. This allows us
to analyze the particulars of dislocation glide in 3+1D (as well as in
arbitrary higher dimensions). In high dimensions, dislocations
are no longer particles but instead extended higher dimensional
p-branes with p=1 in 3+1D and p=2
in 4+1D. We will show that the glide constraint influences the 
{\em center of mass
motion} of the branes while the relative (transversal) motions of these
extended manifolds remain unconstrained. This amounts to a precise 
mathematical description of the `restricted climb' known from metallurgy.
We subsequently turn to the disclination and defect currents and 
presents proofs for deep- but also disappointing traits of these currents:
the mass conservation underlying dislocation glide turns into simple
conservation laws. Another important property of the constraint
is represented in the symmetry of the constrained current. By identifying
this constraint as applying only to 
the unique (rotational symmetry invariant) `singlet' 
component of the dynamical dislocation current, the relation of the glide
constraint to compression is firmly established. In the final section 
(Section (V)), we lift this to the
full non-linear level incorporating the presence of a lattice cut-off
and interstitial matter in the field theory formalism.

\section {Elastic action and defect densities.}

Deformations in ideal crystals are parameterized by displacement fields 
${\bf u} = {\bf R} - {\bf R}_0$, where ${\bf R}$ and  ${\bf R}_0$ refer to 
the actual position and the position in the ideal crystal of the 
constituent,
respectively. The action of an elastic medium contains strain, 
kinetic, and external potential components. These are, in general, 
functionals 
of the displacements ${\bf u}$ 
and their derivatives. 
The kinetic energy density term is simply 
${\cal{K}} = \tfrac 
1 2 \rho ( \partial_t {\bf u})^2$.
The incorporation of the kinetic energy density 
${\cal{K}}$ in the total Lagrangian density leads to the variational 
equations of motion 
for the dynamics of ${\bf u}$ in both 
quantum systems (with bosonic displacements ${\bf u}$) 
and simple classical ones. Much throughout, we will perform a Wick rotation
(wherein the real time $t$ replaced by $-i \tau$) to Euclidean space-time.
In the resulting imaginary-time Euclidean action,
the Lagrangian density now becomes an ``energy''
density given by the {\em sum} of kinetic and potential parts. 
The elastic portion of the 
Lagrangian density follows from the gradient
expansion in terms of the displacements. To leading order 
(`first order elasticity'), the elastic energy density is 
$\partial_i u^a C_{ijab} \partial_j u^b$ with $C_{ijab}$ the 
elastic tensor whose general form is dictated by the symmetries of
the medium. In the remainder of the article,
we will employ Latin indices to refer to spatial components,
while Greek indices will be reserved for 
space-time components. Einstein summation convention
will further be assumed for repeated indices. 
In full generality, the action will further contain
higher order derivatives and anharmonic strain couplings:
\begin {eqnarray}
  S_0 = \int \rmd \tau \rmd {\bf x} ~ {\cal L}_0 \lbrack {\bf u}, 
\partial {\bf u}, \ldots, \partial^{m} {\bf u} \rbrack, 
\label {S0}
\end {eqnarray}
corresponding to the partition function
\begin {eqnarray}
  Z = \int {\cal D} {\bf u} ~ e^{-S_0  \lbrack {\bf u}, \partial {\bf u}, 
\ldots, \partial^{m} {\bf u} \rbrack}. \label {Z0}
\end {eqnarray}

In the absence of singularities, the action above corresponds with the
simple problem of acoustic phonons. However, when defects are present,
the theory becomes far richer. In its
full non-linear form, elasticity 
is closely related to Einstein gravity\cite {KZ}. In
the 1980's, Kleinert  \cite {Kleinert} achieved a considerable progress by his
recognition of the underlying gauge field-theoretic structures,
employing the mathematical machinery of gauge theory to penetrate
deeper into this subject than ever before; much of this is found
in his book. This work focused primarily on the `plasticity' of
3D classical media, and only very recently it was extended to
the problem of quantum melting in 2+1D\cite {ZMN}.

The key notion introduced by Kleinert is the dualization of
the action Eq.(\ref {S0}) into stress variables to recognize
subsequently that stress can be expressed in terms of stress
gauge fields or `stress photons'. Topological defects 
then take the form of {\em sources} for the stress photons.
This follows the same pattern as the vortex (or Abelian Higgs)
duality where the phase modes of the superfluid are dualized
in $U(1)$ gauge fields, mediating the interactions between
vortex sources. This analogy is quite close when only 
dislocations are in the game. The quantum nematics addressed by two of us
earlier\cite {ZMN} are dual superconductors in the same sense that
the quantum disordered superfluids (Mott-insulators) can be
regarded as dual superconductors, the difference being that
in the dislocation condensate shear acquires a Higgs mass 
while in the vortex condensate super-currents acquire
the Higgs mass. When disclinations come into play, we need to
employ rank two tensorial gauge fields
(`double curl gauge fields'). On this level, 
the correspondence with gravity become 
manifest\cite{KZ}. As this construct is not widely known, we 
summarize the basic dualization steps in Appendix A.

\begin{figure}
\includegraphics[width=7cm]{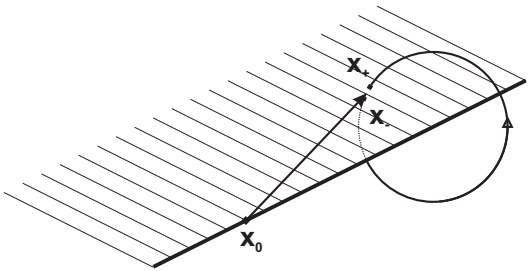}
\caption{Weingarten theorem: encircling a defect line in a
three dimensional medium results in a displacement that depends
exclusively on the starting/ending point ${\bf x}_{\pm}$. The
discontinuity occurs at the Volterra cut (rastered plane).} 
\label{WeingartenFig}
\end{figure}

There is a famous theorem due to  
Weingarten \cite {Weingarten} which states 
that the singular displacement recorded 
by encircling a defect line in a three-dimensional elastic medium may always 
be expressed as a sum of a constant vector and an antisymmetric operator 
acting on the radial vector ${\bf x}$ (see Fig. \ref {WeingartenFig})
\begin {eqnarray}
  \oint_C (\rmd {\bf l} \cdot \nabla) {\bf u} = {\bf u} ({\bf x}_+) - {\bf u} 
({\bf x}_-) =
  {\bf b}_0 + \hat {\mathscr A} ( {\bf x} - {\bf x}_0). 
\label {Weingartenstrain}
\end {eqnarray}

The reference point ${\bf x}_0$ can be arbitrary, but  when fixed
on the defect line it defines the Burgers vector ${\bf b}$ or 
`dislocation charge' of the defect. The three-dimensional
antisymmetric operator $\hat {\mathscr A}$ can be represented as the cross
product of the radial vector ${\bf x}$  with the pseudo-vector $\Omega$:
$\hat {\mathscr A} {\bf x} = \Omega \times {\bf x}$. The pseudo-vector 
$\Omega$ is
the Frank vector or the `disclination charge' of the defect. This theorem makes
immediately clear that dislocation and disclination charges are related. For 
instance, a dislocation can be viewed as a bound state of a disclination
and an anti-disclination, while the disclination corresponds with a
`stack' of an infinite number of dislocations. When addressing
linear elasticity, the proper topological
current turns out to be a particular combination of dislocation- and
disclination currents (the `topological defect current'\cite{Kleinert}).

Employing the dualization recipe of Appendix A in three classical
dimensions, we can straightforwardly derive 
differential expressions for the dislocation and disclination densities
respectively,
\begin {eqnarray}
  \alpha_i^a &=& \varepsilon_{ijk} \partial_j \partial_k u^a, 
\label {staticdislocation} \\
  \Theta_i^a &=& \tfrac 1 2 \varepsilon_{ijk} \varepsilon_{abc} 
\partial_j \partial_k \partial_b u^c, \label {staticdisclination}
\end {eqnarray}
where $\varepsilon$ denotes the 3D antisymmetric (Levi-Civita) tensor.
For a defect line in a given spatial direction $i$, the charges in 
Eqs.(\ref{staticdislocation}, \ref {staticdisclination}),
derived in Appendix A via a dualization procedure,
coincide with the Burgers and Frank vectors, respectively.
To see this, consider, for instance, the edge dislocation
of Fig.(\ref{Glidefig}). Here, in going around the dislocation
point there is a net change of the displacement fields.
Couched in standard mathematical terms, the Burgers vector signaling
the total variation in the displacement field is, as well known,
\begin {eqnarray}
{\bf b} = b \hat{e}_{x} = \oint {\bf \nabla} u^{x} \cdot {\bf dl}
= \int \int_{{\cal{D}}} {\bf \nabla} \times {\bf \nabla} u^{x} \cdot 
\bf{dA}.
\label{Burgers}
\end {eqnarray} 
In Eq.(\ref{Burgers}), 
$\hat{e}_{x}$ denotes a unit vector along the horizontal direction
and $\bf{dA}$ is a planar area element within a region
${\cal{D}}$ containing the dislocation point. We recognize the argument
of the last integrand in Eq.(\ref{Burgers}) as 
the dislocation density of Eq.(\ref{staticdislocation})-
${\bf \nabla} \times {\bf \nabla} u^{x} = \alpha^{x}_{j} \hat{e}_{j}$.
Similar considerations lead to the identification of 
Eq.(\ref{staticdisclination}) as the angular mismatch
(the disclination) density. A nice construct for its visualization
is the Volterra cut \cite{Kleinert}. Needless to say, the derivation of
the defect densities via integrals such as that in 
Eq.(\ref{Burgers}) is {\em topological}; these forms are
valid for any crystal regardless of its fundamental
constituents (whether they are classical or quantum 
matter of one particular statistics or another).
In what will briefly follow, within the arena
of dynamical defects (Section III),
we will elevate the densities of
Eqs.(\ref{staticdislocation}, \ref{staticdisclination})
into the zero (temporal) component of Euclidean space-time
defect currents. As a curiosity, an old famous anecdote 
concerning time 
like defects is provided in \cite{commentBurgerstau}. 

As stated earlier, dislocations and disclinations are
not independent entities: a dislocation may be viewed as a
disclination- antidisclination pair
while a disclination is an infinite stack 
of dislocation lines. Thus, the densities
of Eqs.(\ref{staticdislocation}, \ref{staticdisclination})
harbor redundant information. This redundancy may be removed
by fusing the dislocation and disclination densities 
into a more fundamental (double curl) 
topological defect density,
\begin {eqnarray}
  \eta_i^a &=& \tfrac 1 2 \varepsilon_{ijk} \varepsilon_{abc} \partial_j 
\partial_b \left (
  \partial_k u^c + \partial_c u^k \right ) = \nonumber \\
  &=& \Theta_i^a + \tfrac 1 2 \partial_m \left \lbrack \varepsilon_{min} 
\alpha_a^n +
  \varepsilon_{man} \alpha_i^n - \varepsilon_{ian} \alpha_m^n 
\right \rbrack. \label {eta3d}
\end {eqnarray}

Within this publication, most of our studies of these densities 
will be within the confines of the linearized theory of elasticity.
When higher derivatives of the displacement field ${\bf u}$
appear in the action Eq.(\ref {S0}), the defect densities discussed above
will fail to capture all of the relevant singularities.
The role of higher order corrections will be briefly touched upon in section V.

Depending on the alignment of the charge vector and the three dimensional 
defect line, dislocations are classified as edge (perpendicular, $i \neq a$) 
or screw (parallel, $i = a$), while the disclinations may be of the wedge 
(parallel, $i = a$), sway (perpendicular, $i \neq a$) or twist
variants.

\section {Dynamical dislocation- and disclination currents in arbitrary
dimensions.}

In this section, we will generalize the gauge theoretical description
of defect currents to higher dimensions limiting ourself to the
linearized level (first order elasticity).

As a defect drifts through the medium, the defect charge 
transforms into dynamical defect current. For the
particular case of the 2+1-dimensional medium, such issues 
were addressed for the first time by two of us \cite {ZMN}. As a first
extension, let us now consider how these currents look like
in 3+1D. One has now to  dualize the full action including
the kinetic term. This is a straightforward extension of the
2+1D case,
\begin {eqnarray}
  J_{\mu \nu}^a &=& \varepsilon_{\mu \nu \lambda \rho} \partial_\lambda \partial_\rho u^a,
\label {currents3J} \\
  T_{\mu \nu}^{\alpha \beta} &=& \varepsilon_{\mu \nu \lambda \kappa} \partial_\lambda \partial_\kappa \omega^{\alpha \beta},
\label {currents3T}
\end {eqnarray}
with the local rotation given by
\begin {eqnarray}
  \omega^{\alpha \beta} = \tfrac 1 2 \varepsilon_{\alpha \beta \rho \sigma} \partial_\rho u^\sigma. \label {omegaab}
\end {eqnarray}
The currents are now two forms in the `lower' space time labels, actually
keeping track of the fact that the defect lines spread out in world sheets
(or 2-Branes) in 3+1D space-time. Obviously, by taking one of these labels
to correspond with the time direction one immediately recovers the static
densities, $J_{\tau i}^a = \alpha_i^a$ and $T_{\tau i}^{\tau a} = \Theta_i^a$.
The currents with two lower spatial indices correspond with the dynamical
currents: the lower indices $ij$ refer to a current in the $i$ direction,
of a defect line extending into the $j$ direction.

\begin{figure}
\includegraphics[width=7cm]{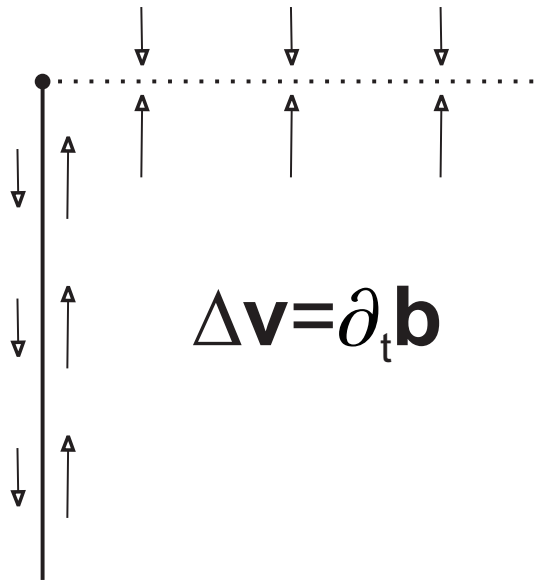}
\caption{Discontinuity in the velocity field or the `velocity dislocation':
such a discontinuity, though physically possible, violates the assumptions
of the Weingarten theorem Eq.(\ref {Weingartenstrain}). Depending 
on the orientation of the
Volterra cut with respect to the `velocity Burgers vector' it represents
slip of surfaces (parallel, full line) or adding/removing material
(perpendicular, dashed line). For simplicity, we use the word `slip' for
both motions.} \label {Figslip}
\end{figure}

In solids (and other condensed matter systems), 
the medium cannot displace in the
temporal direction $\tau$: 
$u_{\tau} = 0$. Consequently, in the 
definition of Eq.(\ref {currents3J}),
the`upper' Burgers-labels $^{a}$ of the dislocation
currents $J_\mu^a$ are purely spatial. Formally, these spatial directions 
can be regarded as `flavor' degrees of freedom \cite {commentBurgerstau}.
We derive Eq.(\ref {currents3T}) by employing a generalization
the static current of Eq.(\ref {staticdisclination}) along with 
an application of the Weingarten theorem (Eq.(\ref {Weingartenstrain})). 
The local rotation operator Eq.(\ref {omegaab})
used in the definition has also been generalized to deal with the
in-equivalence of space and time.
When one of the indices $\alpha, \beta$  
is temporal, Eq.(\ref {omegaab}) reduces to the usual\cite{Kleinert} spatial
rotation $\omega^{\tau i} = \omega^i$. The corresponding disclination current
then carries the Frank charge of the defect. 
What is the physical meaning of 
the operator Eq.(\ref {omegaab}) when both indices $\alpha, \beta$ 
are spatial? Due to the constraint on temporal
displacements ($u^\tau = 0$), the only label that can take the temporal value
$\tau$ is $\rho$ and the generalized rotation represents the velocity field
\begin {eqnarray}
  \omega^{ab} = \tfrac 1 2 \varepsilon_{abc} v^c. \label {omegavelocity}
\end {eqnarray}

The corresponding dynamical disclination current $T_{\mu \nu}^{ab}$ records
discontinuity, a `dislocation', in the velocity field. Physically, 
this discontinuity represents slip of two surfaces relative to each other, 
acting  as a `defect factory' in the crystal (Fig. \ref {Figslip}). However, 
the Weingarten theorem, which is the underneath the definitions
Eq.(\ref {currents3J} - \ref {currents3T}), assumes that all 
the symmetrized 
strains $u_{\rho, \sigma} = \tfrac 1 2 (\partial_\rho u^\sigma 
+ \partial_\sigma u^\rho)$ are smooth everywhere. Choosing one of the 
indices $\rho$ or $\sigma$ to correspond with time and the other to space, 
this strain becomes a local rotation,
\begin {eqnarray}
  u_{\tau, a} = \tfrac 1 2 v^a = \varepsilon_{abc} \omega^{bc}.
\end {eqnarray}
A nonzero value of the disclination current $T_{\mu \nu}^{ab}$ would mean 
discontinuity in the velocity field, which in turn violates the Weingarten 
theorem. To avoid confusion between real disclination densities and these 
surface slips we redefine them as follows,
\begin {eqnarray}
  T_{\mu \nu}^a = \tfrac 1 2 
\varepsilon_{\mu \nu \lambda \kappa} \varepsilon_{a b c}
  \partial_\lambda \partial_\kappa \partial_b u^c, \label {currents3Ta}
\end {eqnarray}
so that only the Frank vector is represented by the upper label 
(disclination `flavor'). When generalized to 3+1D, the dualization procedure 
summarized in appendix A indicates that the surface 
slips which we left out will eventually 
play no role, because they do not couple to stress photons.

Let us now derive the form of the topological currents in
arbitrary dimensions. The key lies, once again, in the Weingarten theorem 
of Eq.(\ref {Weingartenstrain}) which
is valid in any dimension. In $D$ dimensional space, 
the Burgers vector is a 
$D$-dimensional while the 
disclination charge is 
a tensor of rank $D$-2. The rank of the 
disclination charge may be ascertained from the fact that the antisymmetric 
tensor $\hat {\mathscr A}$ can be written as a contraction of Levi-Civita 
symbol and another, $D$-2 dimensional  antisymmetric tensor:
 ${\mathscr A}_{ij} = \varepsilon_{i j k_1 k_2 \ldots k_{D-2}} 
\Omega_{k_1 k_2 \ldots k_{D-2}}$. Thus, two, three and four dimensional 
disclinations are characterized by a Frank scalar, vector and antisymmetric 
rank 2 tensor respectively. The $D$+1-dimensional extension of the definitions 
of the currents Eq.(\ref {currents3J},\ref {currents3Ta}) are
\begin {eqnarray}
  J_{\mu_1 \ldots \mu_{D-1}}^a &=& \varepsilon_{\mu_1 \ldots \mu_{D-1} \nu \lambda} \partial_\nu \partial_\lambda u^a, 
\label {currentJd} \\
  T_{\mu_1 \ldots \mu_{D-1}}^{a_1 \ldots a_{D-2}} &=& \tfrac 1 2 \varepsilon_{\mu_1 \ldots \mu_{D-1} \nu \lambda} \varepsilon_{a_1 \ldots a_{D-2} b c} \partial_\nu \partial_\lambda \partial_b u^c. \label {currentTd}
\end {eqnarray}

Both of the currents of Eqs.(\ref{currentJd},\ref{currentTd}) 
are antisymmetric in the lower indices, and represent 
oriented $p=D-1$-branes. The disclination currents are antisymmetric 
also in their upper indices where the Frank charge is displayed.
We reiterate that within linear elasticity the true (double curl) topological
defect density of Eq.(\ref{etad})
removes the redundant information given by the inter-related
dislocation and disclination currents. Extending
the duality procedure of linearized
elasticity (appendix A) to arbitrary dimensions, we easily deduce that
the fundamental double curl defect current within
a $D$+1 dimensional medium is
\begin {eqnarray}
  \eta_{\eta_1 \eta_2 \ldots \eta_{D-1}}^{\nu_1 \nu_2 \ldots \nu_{D-1}} = \nonumber \\
  \tfrac 1 2 \varepsilon_{\eta_1 \ldots \eta_{D-1} \kappa \lambda}
  \varepsilon_{\nu_1 \ldots \nu_{D-1} \rho \sigma}
  \partial_\kappa \partial_\rho (\partial_\lambda u^\sigma + \partial_\sigma u^\kappa). \label {etad}
\end {eqnarray}

This current is antisymmetric under exchange of any two lower or any two 
upper indices. This defect current is 
related to the dislocation and disclination 
currents Eq.(\ref {currentJd} - \ref {currentTd}) via
\begin {eqnarray}
  \eta_{\eta_1 \ldots \eta_{D-1}}^{\nu_1 \ldots \nu_{D-1}} = \sum_{i=1}^{D-1} (-)^{i+1} \delta_{\tau \nu_i}
  T_{\mu_1 \ldots \mu_{D-1}}^{\nu_1 \ldots \nu_{i-1} \nu_{i+1} \ldots \nu_{D-1}} + \nonumber \\
  + \tfrac 1 2 \varepsilon_{\mu_1 \ldots \mu_{D-1} \kappa \lambda} \partial_\kappa J_{\nu_1 \ldots \nu_{D-1}}^\lambda +
  \tfrac 1{2 (D-2)!} \varepsilon_{\mu_1 \ldots \mu_{D-1} \kappa \lambda} \nonumber \\
  \times \varepsilon_{\nu_1 \ldots \nu_{D-1} \rho \sigma}
  \varepsilon_{\rho \lambda \xi_1 \ldots \xi_{D-2}} \partial_\kappa J_{\xi_1 \ldots \xi_{D-2}}^\sigma. \label {generalJT2eta}
\end {eqnarray}

To avoid surface slips, one of the upper indices in the definition 
Eq.(\ref {etad}) must be temporal when addressing the disclination currents 
$T$. When all the upper indices are spatial, such components record only the 
derivatives of the dislocation currents
of Eq.(\ref{currentJd}).

\section {The electrically charged medium and glide.}

Let us now switch gear, to derive mathematical expressions of the glide 
constraint in terms of the currents introduced above. 
We consider an electrically charged medium (`Bosonic Wigner
crystal'), both because it is interesting on its own right \cite {Wigner,
Ceperley, EMZMNC}, and also because
it provides us with a convenient vehicle for the derivation
of general expressions for the
glide constraint. Later on, we will independently
derive these glide constraints by direct {\em mass conservation} 
without resorting to local gauge invariance to implement it.
As mass conservation pertains to a scalar quantity, 
such a conservation law translates into a condition on a
linear combination of the topological 
defect currents which is necessarily
invariant under spatial rotations. 
Conservation laws (equivalent to gauge invariance)
may greatly restrict the system dynamics leading to 
an effective reduction in the dimensionality. We will now illustrate 
how this indeed transpires in solids: in linear elasticity,
mass (`charge') conservation allows only a glide motion of a dislocation.

We proceed with the treatment of a charged uniformly
charged medium by employing the standard 
electromagnetic (EM) gauge field formulation.
As usual, in Euclidean space, the electrical currents defined in terms
of the constituent particles are minimally coupled to the electromagnetic
potentials via  $i j_\mu A_\mu$.  The spatial current components 
relate to the velocity of 
the charged particles, so that every particle with charge $e$ contributes 
to the current as $(e ~ \partial_\tau u^i)$. For a medium with $n_e$ of
such particles per unit volume, the corresponding current density will be 
$j_i = (n_e e) ~ \partial_\tau u^i$. The time components describe the 
coupling between the Coulomb potential $A_\tau$ and the charge density 
$j_\tau$. For sufficiently small strains, the density is 
$[(n_e e) ~ (1 - \partial_i u^i)]$, subtracting the constant contribution 
(compensating electrical background). On this linearized level, we
should add the following term to the Lagrangian of quantum elasticity
to describe the coupling of the EM field to the electrically charged medium, 
\begin {eqnarray}
  {\cal L}_{EM} = i (n_e e) ~ 
\lbrack A_i \partial_\tau u^i - A_\tau \partial_i u^i \rbrack. 
\label {LEM}
\end {eqnarray}

The electromagnetic fields further have their 
own dynamics, described by the Maxwell term ${\cal L}_{Maxwell} = 
\frac 1 4 F_{\mu \nu} F_{\mu \nu}$ with field strengths 
$F_{\mu \nu} = \partial_\mu A_\nu - \partial_\nu A_\mu$.

Eq.(\ref{LEM})  must be invariant under EM gauge transformations 
$A_\mu \to A_\mu + \partial_\mu \alpha(x^\nu)$, with $\alpha (x^\nu)$ being
an arbitrary, non-singular scalar function. The Maxwell term is automatically 
gauge invariant but the demand of gauge-invariance of  the minimal coupling 
term Eq.(\ref{LEM}) implies that the currents are locally conserved. After
partial integration,
\begin {eqnarray}
  \delta {\cal L}_{EM} =
i j_\mu \partial_\mu \alpha \longrightarrow - i \alpha \partial_\mu j_\mu.
\label {emmin}
\end {eqnarray}

As $\alpha (x^\mu)$ is arbitrary, it follows that the current is conserved,
$\partial_\mu j_\mu = 0$ -- the standard result that gauge invariance implies
the conservation of the gauge charge. Let us now see how this works out
considering instead the displacement. Performing a gauge 
transformation on Eq.(\ref{LEM}) one finds immediately a gauge
non invariant part, 
\begin {eqnarray}
  \delta {\cal L}_{EM} & = & i (n_e e) ~ \lbrack \partial_i \alpha 
\partial_\tau u^i - \partial_\tau \alpha \partial_i u^i \rbrack  
\nonumber \\
  & \longrightarrow & i (n_e e) \alpha (x^\mu) ~ \lbrack 
\partial_\tau \partial_i u^i - \partial_i \partial_\tau u^i \rbrack,
\label {glidestrain}
\end {eqnarray}
and gauge invariance implies that
\begin {eqnarray}
\lbrack  \partial_\tau \partial_i u^i - \partial_i \partial_\tau u^i 
\rbrack = 0
\label{keyresult}
\end {eqnarray}

Although this derivation is very simple, this equation is a key result of
this paper. Combining it with definition of the dislocation current in
arbitrary dimensions Eq.(\ref {currentJd}) and using the contraction identity,
\begin {eqnarray}
  \varepsilon_{\tau a \mu_1 \ldots \mu_{D-1}} 
\varepsilon_{\mu_1 \ldots \mu_{D-1} \nu \lambda} = (D - 1)! 
~ \left | \begin {array}{cc} \delta_{\tau \nu} & \delta_{\tau \lambda} \\ \delta_{a \nu} & \delta_{a \lambda} \end {array} \right |,
\end {eqnarray}
it follows that the requirement of conservation of electrical charge
has transformed itself into a constraint acting on the dislocation current,
\begin {eqnarray}
\varepsilon_{\tau a i_1 \ldots i_{D-1}} J_{i_1 \ldots i_{D-1}}^a = 0. 
\label {glidecurrent}
\end {eqnarray}

This is none other than the glide constraint acting on the dislocation
current in arbitrary dimensions! 

At first sight this might appear as magic but it is easy to see
what is behind this derivation. In order to derive
Eq.(\ref {LEM}) from Eq.(\ref {emmin}) one has to assume
that the gradient expansion is well behaved, i.e. the displacements
should be finite. This is not the case when interstitials are present,
because an interstitial is by definition an object which can dwell away
an infinite distance from its lattice position. Hence, in the starting
point Eq.(\ref{LEM}) it is implicitly assumed that the interstitial
density is identical zero. The gauge argument then shows that gauge 
invariance exclusively communicates with the {\em non-integrability}
of the displacement fields, Eq.(\ref{keyresult}). These non-integrabilities
are of course nothing else than the topological currents -- the glide
constraint is a constraint on the dislocation current. If the glide
constraint was not satisfied, electrical conservation would be
violated locally, i.e. electrical charges would (dis)appear spontaneously,
as if the dislocation is capable to create or destroy crystalline
matter. In the absence of interstitials this is not possible and,
henceforth, dislocations can only glide. The key is, of course, that,
by default, dislocation currents are decoupled from compressional stress in
the linear non-relativistic theory.

We may, indeed, equivalently 
derive Eq.(\ref{glidecurrent}) without explicitly 
invoking EM gauge invariance to arrive at Eq.(\ref{keyresult}).
Instead, we may directly rely from the very start on {\em mass conservation}-
the continuity equation of the mass currents
(which, as alluded to above, is 
equivalent to local gauge invariance), \cite{explaintau}
\begin {eqnarray}
  \partial_\tau \rho + \nabla \cdot {\bf j} = 0. \label {gliderho}
\end {eqnarray}

To see how this is done directly, we compute the 
various mass current components $j_{\mu}$. 
By simple geometrical considerations, within
the linear elastic regime, the 
mass density \begin {eqnarray}
\rho =  \rho_{0}[ 1 - \partial_{i} u^{i}],
\label{j0.}
\end{eqnarray}
with $\rho_{0}$ the uniform background 
value: the divergence of ${\bf{u}}$ 
(signaling the local volume increase)
yields the negative net 
mass (`charge') density variation at any point.
Similarly, the spatial current density  
\begin {eqnarray}
{\bf j} = \rho_{0} \partial_{\tau} {\bf u}.
\label{ji.}
\end{eqnarray}
Compounding the mass continuity equation of Eq.(\ref{gliderho})
with the physical identification of the current 
(Eqs.(\ref{j0.}, \ref{ji.})),
we obtain Eq.(\ref{keyresult}) from
which Eq.(\ref {glidecurrent}) follows.
In section (\ref{GCLCA}), we will return 
to such a physical interpretation
of the glide constraint
from this perspective in order
to determine corrections to the 
glide principle which follow
from anharmonic terms. We emphasize that the mass conservation
law leading to Eq.(\ref{keyresult}) trivially holds in any medium
regardless of the underlying statistics \cite{camino}  
of potential quantum systems or their dimension. 
Furthermore, within the linear elastic
regime, such a discussion
highlights the validity of Eq.(\ref{keyresult}) and the ensuing
glide equation of Eq.(\ref{glidecurrent}) 
(when interpreted as density matrix averages) in crystals 
at any temperature in which strict linear order mass conservation 
condition is imposed on all configurations. Needless
to say, as temperature is elevated, 
a departure occurs from such an imposed linear order condition
through the enhanced appearance and diffusion of interstitials and 
vacancies leading to climb motions. 
The restriction on the dynamics in this regime 
is captured by a higher order 
variant of Eq.(\ref{glidecurrent}) (Eq.(\ref {glidecubic})) 
which will be derived later on.

Let us now pause to consider
what Eq.(\ref {glidecurrent}) means physically. In two
spatial dimensions, Eq.(\ref {glidecurrent})
implies that the dislocation currents have to
be symmetric \cite {ZMN}: $\varepsilon_{ab} J_a^b = J_x^y - J_y^x = 0$. 
Now, consider Fig. (1). Here, the 
Burgers vector is pointing in the horizontal x-direction
implying that $J^y_{\mu} =0$, while the glide constraint
reduces to $J^x_y = 0$. This current ($J^x_y$) 
is, by its very definition, the climb current perpendicular
to the Burgers vector.

In three and higher spatial dimensions, the story is less easy, the reason 
being that the constraint on the motion is less absolute. This is of course
known in the classic theory \cite {Nabarro, Kleinert}, but the reader might 
convince him/herself
that making use of Eq. (\ref {glidecurrent}) the analysis is much
helped as compared to the rather pain staking effort based on the
`intuitive' arguments. In 3+1D, the constraint of 
Eq.(\ref {glidecurrent}) becomes
\begin {eqnarray}
  \epsilon_{\tau a b c} J^a_{bc} = 0. \label{glide3D}
\end {eqnarray}

\begin{figure}
\includegraphics[width=7cm]{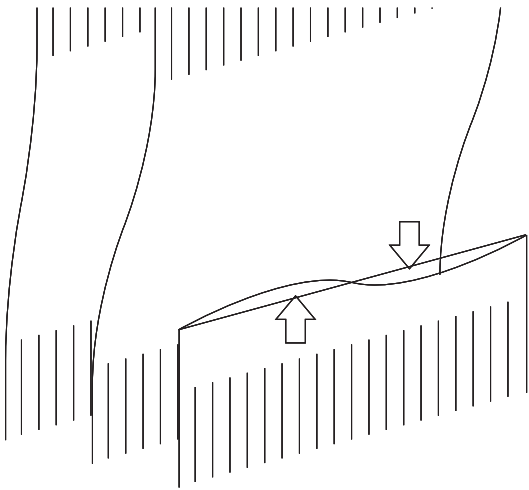}
\caption{Restricted climb in the 3+1D medium: although the dislocation line
cannot displace as a whole in the direction perpendicular to both itself  and 
its  Burgers
vector (climb direction), local segments may climb at the  expense of their
neighbors' height.} \label {Fig3Dglide}
\end{figure}

Let us first consider a screw dislocation. These
correspond with dislocation currents of the form $J^a_{a\mu}$ (i.e. the 
static $\mu = \tau$ component corresponds with the orientation of the
dislocation loop being parallel to the Burgers vector). It follows
immediately that the constraint Eq.(\ref {glide3D}) is not acting 
on screw dislocations and, henceforth, screw dislocations can
move freely in all directions. Edge dislocations are the other
extreme, corresponding with dynamical currents of the form 
$J^a_{b\mu}$ where $a$ and $b$ are orthogonal. The condition $\mu = a$
corresponds with glide: an edge dislocation with its loop oriented
in a direction ($b$) perpendicular to the Burgers vector ($a$) can still 
move freely in the direction of the Burgers vector ($a$). The displacement
of a dislocation along the line is not a topological object and the current
with two identical lower indices ($\mu = b$) vanishes. The glide constraint 
only strikes when all three labels are different. Let us consider a 
dislocation line extending in $z$ direction with Burgers vector in the $x$ 
direction (see Fig. \ref {Fig3Dglide}). The only nonzero components of the
current are $J_{z\mu}^x$ and the glide constraint becomes $J^x_{zy} = 0$.
This automatically forbids any motion in the $y$ direction, that is 
perpendicular both to the dislocation line and its Burgers vector. The 
constraint is `leaky' due to the extended nature of the defect. The 
material 
needed for the climb of one segment of the dislocation can be 
supplied by an adjacent segment. The glide constraint has therefore only
a real meaning through its integral form,
\begin {eqnarray}
  0 = \int dV \varepsilon_{\tau abc} J_{ab}^c. \label {glide3dintegral}
\end {eqnarray}

As illustrated in Fig. \ref {Fig3Dglide}, only the dislocation's
`center of mass' is prohibited to  move in the climb direction. Local 
segments of the line may still move at expense of their neighbors, 
effectively transporting matter along the defect line. The `leaky',
locally defined, constraint of Eq.(\ref {glide3dintegral}) corresponds
with the intuitive idea of `restricted climb' found in the elasticity 
literature.

Having a mathematical definition of the glide constraint at our 
fingertips enables us to address its incarnations in even higher dimensional
systems where direct visualization is of little use. 
Notwithstanding that such crystals are, of course, 
not to be found in standard condensed matter, 
higher dimensional 
glide constraints might have implications for `emergence theories'  of 
fundamental phenomena resting on  elasticity 
theory \cite {KZ, KleinertBrJP}. 
Let us for example look how the constraint Eq.(\ref {glidecurrent}) acts in 
a 4+1D crystal. A dislocation is now a 2-brane, say, a  plane extending in 
$y$ and $z$ directions. When its Burgers vector lies in
this plane, this brane is analogous to a screw dislocation in the sense 
that its motions are  not affected by the glide constraint. The other extreme
is the `edge dislocation brane' with a  Burgers vector perpendicular to the 
brane, say in the $x$ direction. In this case  the nontrivial currents 
correspond with $J_{y z \mu}^x$. As noted earlier
for three-dimensional defects, topological currents do not record any 
motion taking place within the defect-brane but only in directions
perpendicular to it. The bottom line is that besides the static current
(density) $J_{yz\tau}^x$, the  topological dynamical currents are $J_{yzx}^x$
and $J_{yzw}^x$,  representing dislocation glide and climb respectively. 
The glide constraint Eq. (\ref {glidecurrent}) forbids the latter `in
integral form', allowing climb of a certain brane element only at the
expense of the brane volume taken by a neighboring element.

\subsection {The glide constraint and motions of general defects.}

Up to this point we focused on dislocations, the traditional scenery
for glide motions. However, as alluded to in section II, dislocation
and disclination currents are not independent.
The truly fundamental topological currents  
are the (double curl) defect currents of Eq.(\ref{etad}). 
A natural question to 
ask is what the form of the glide constraint is on these fundamental
entities. This task is made easy by Eq.(\ref{keyresult}).
In what follows, we will
derive new generalized glide equations for a medium in
which both dislocations and disclinations are present.
  
Let us first consider the
most trivial case, the 2+1D crystal. Excluding surface slips,
Eq.(\ref {generalJT2eta}) becomes
\begin {eqnarray}
  \eta_\mu^\nu = \delta_{\nu \tau} T_\mu - \tfrac 1 2 \partial_\lambda
\left \lbrack
  \varepsilon_{\lambda \mu a} J_\nu^a + \varepsilon_{\lambda \nu a} J_\mu^a -
  \varepsilon_{\mu \nu a} J_\lambda^a \right \rbrack. \label {eta2d1}
\end {eqnarray}

In two spatial dimensions, we immediately note from 
Eq.(\ref {eta2d1}) that the glide constraint implies 
that $\eta_i^i = 0$.

The extension to arbitrary dimensions is straightforward. 
We start with the definition Eq.(\ref {etad})
and then employ the Levi-Civita contraction identity to obtain
\begin {eqnarray}
  \varepsilon_{\mu a_1 \ldots a_{D-1} \rho \lambda} \varepsilon_{\tau a_1 \ldots a_{D-1} b c} 
  = (D-1)! \left | \begin {array}{c c c} \delta_{\mu \tau} & \delta_{\mu b} & \delta_{\mu c} \\ \delta_{\rho \tau} & \delta_{\rho b} & \delta_{\rho c} \\ \delta_{\lambda \tau} & \delta_{\lambda b} & \delta_{\lambda c} \end {array} \right |. \label {LeviCivitaD2}
\end {eqnarray}
The terms containing $u^\tau$ vanish. The no-slip condition 
($\partial_\mu \partial_\nu v^a  = \partial_\nu \partial_\mu v^a$) 
needs to be imposed supplanted by the glide constraint in its strain 
form of Eq.(\ref {glidestrain}). The outcome is that the scalar spherical 
tensor component of the general defect current 
tensor of Eq. (\ref {generalJT2eta}) vanishes
\begin {eqnarray}
  \eta_{a_1 a_2 \ldots a_{D-1}}^{a_1 a_2 \ldots a_{D-1}} = 0. \label {glideeta}
\end {eqnarray}

We next investigate what the glide constraint implies for
disclinations. The constraint of Eq.(\ref {glideeta}) 
does not shed much light on this issue. Linear elasticity directly 
addresses dislocations: as we already argued, 
one of the upper indices in the defect current of
Eq.(\ref {etad}) needs to be temporal in order to 
record disclination currents. 
On the other hand, it is clear that disclinations somehow do know about glide
because they can be viewed  as an infinite stack of dislocations. As an 
example, consider a wedge disclination: the material
added by the Volterra cut is proportional to the Frank charge and this
should not change over time. It turns out that the matter conservation 
associated  with  a general defect distribution is related to the
Volterra cutting procedure. After the cut has been applied, no additional 
material should be introduced, which is the same as the requirement that
all symmetrized strains and  derivatives thereof are
smooth everywhere, including the locus of the Volterra cut.
This condition is hard wired into the proof of the generalized 
Weingarten theorem Eq.(\ref {Weingartenstrain}). As we will now show, 
the ramification of this principle for disclinations is a conservation law. 

To make headway, it is convenient to first consider
Euclidean Lorentz-invariant $D+1$ space time. The non-relativistic
case will turn out to be a special case which directly follows from
imposing the condition of the absence of time like displacements 
($u^\tau = 0$). What follows rests heavily on elastic analogs of 
identities in differential geometry which are discussed in detail
in the last part of Kleinert's book\cite{Kleinert}. 
First, we introduce the tensor
\begin {eqnarray}
  R_{\mu \nu, \rho \sigma} = (\partial_\mu \partial_\nu - \partial_\nu \partial_\mu) \partial_\rho u^\sigma
\end {eqnarray}
representing the Riemann-Christoffel curvature tensor in the geometrical
formulation of the theory of elasticity (the $u$'s are crystal displacements). 
The smoothness assumptions underlying the Weingarten theorem may 
be expressed as,
\begin {eqnarray}
  0 &=& (\partial_\mu \partial_\nu - \partial_\nu \partial_\mu) (\partial_\rho u^\sigma + \partial_\sigma u^\rho), \label {smoothstrain} \\
  0 &=& (\partial_\mu \partial_\nu - \partial_\nu \partial_\mu) \partial_\lambda (\partial_\rho u^\sigma + \partial_\sigma u^\rho), \label {smootherstrain}
\end {eqnarray}
which in turn imply  smoothness of the displacement second derivative 
\begin {eqnarray}
  0 = (\partial_\mu \partial_\nu - \partial_\nu \partial_\mu) \partial_\lambda \partial_\rho u^\sigma. \label {verysmooth}
\end {eqnarray}
Cycling through the indices $\mu$, $\nu$, and $\lambda$ of the smoothness 
equation Eq.(\ref {verysmooth}), we obtain the Bianchi identity for the 
displacement fields,
\begin {eqnarray}
  \partial_\mu R_{\nu \lambda, \rho \sigma} + \partial_\nu R_{\lambda \mu, \rho \sigma} + \partial_\lambda R_{\mu \nu, \rho \sigma} = 0. \label {Bianchi}
\end {eqnarray}

Starting from the other end, let us analyze a candidate for a
disclination conservation law, corresponding with
$\partial_\mu T_{\mu \nu_1 \nu_2 \ldots \nu_{D-2}}^{\phantom {\mu} \alpha_1 
\alpha_2 \ldots \alpha_{D-1}}$. This can be expressed in terms of the 
Riemann tensor where ${\alpha}$ refers to a string of indices labeled
by ${\alpha}$,
\begin {eqnarray}
  \partial_\mu T_{\mu \lbrace \nu \rbrace}^{\phantom {\mu} \lbrace \alpha \rbrace} &=&
  \tfrac 1 2 \varepsilon_{\mu \lbrace \nu \rbrace \kappa \lambda} \varepsilon_{\lbrace \alpha \rbrace \rho \sigma}
  \partial_\mu \partial_\kappa \partial_\lambda \partial_\rho u^\sigma \nonumber \\
  &=& \tfrac 1 2 \varepsilon_{\mu \lbrace \nu \rbrace \kappa \lambda} \varepsilon_{\lbrace \alpha \rbrace \rho \sigma}
  \partial_\mu (\partial_\kappa \partial_\lambda - \partial_\lambda \partial_\kappa + \partial_\lambda \partial_\kappa) \partial_\rho u^\sigma \nonumber \\
  &=& \tfrac 1 2 \varepsilon_{\mu \lbrace \nu \rbrace \kappa \lambda} \varepsilon_{\lbrace \alpha \rbrace \rho \sigma}
  \partial_\mu R_{\kappa \lambda, \rho \sigma} - \partial_\mu T_{\mu \lbrace \nu \rbrace}^{\phantom {\mu} \lbrace \alpha \rbrace}.
\end {eqnarray}

The first term is zero as a consequence of the contraction of the 
Bianchi identity Eq.(\ref {Bianchi})
and the Levi-Civita symbol $\varepsilon_{\mu \lbrace \nu \rbrace \kappa 
\lambda}$. This implies that relativistic disclination currents are 
conserved. The non-relativistic case is just a special case: the 
vanishing of time like displacements means that all upper labels are
space-like ($\alpha \rightarrow a$) and it follows, 

\begin {eqnarray}
  \partial_\mu T_{\mu \nu_1 \nu_2 \ldots \nu_{D-2}}^{\phantom {\mu} a_1 a_2 \ldots a_{D-2}} = 0. \label {conservationT}
\end {eqnarray}

In the 2+1D medium, only wedge dislocations exist and the message of the 
conservation law Eq.(\ref {conservationT}) is clear: the Frank charge-
the angle of an inserted wedge in the Volterra construction- 
behaves as a trivially conserved scalar component of a tensor. 
In higher dimensions, the disclination current similarly
behaves as a conserved tensorial current. The defect density 
has information regarding both
dislocations and disclinations. The disclination conservation
law has separate ramifications for the defect density.
The fundamental condition  Eq.(\ref {verysmooth}) can be directly
rewritten into a conservation law for the defect density of a similar
form as for the disclinations,

\begin {eqnarray}
  \partial_\mu \eta_{\mu \nu_1 \ldots \nu_{D-2}}^{\alpha_1 \alpha_2 \ldots \alpha_{D-1}} = 0.
\label{conseta}
\end {eqnarray}

This is not surprising as defect currents are proportional to
to disclination currents.

This completes the picture: the conservation of matter mandates that
the `proper' disclinations currents are also conserved. However,
the `handicapped' dislocation currents are not conserved a-priori 
(disclinations form their sources) but they have to pay the price that
they can only glide.

\subsection {Symmetry properties of the constraint}

The static dislocations and disclinations of higher dimensional media are
geometrically complex entities such as lines, sheets or $d$-branes 
with Burgers vectors and Frank tensors attached. When in motion, these 
branes sweep the additional time dimension. For instance, defects in two
space dimensions  are point-like particles turning into world-lines in
space-time, in three space dimensions they form loops spreading out in
strings etc. Nevertheless, regardless of the embedding dimensionality, 
all of these `branes' share an universal property: there is a unique
direction perpendicular to the brane. 
This direction can be related to the dynamical defect currents by contracting
it with the $D-1$ dimensional antisymmetric tensor having one index set
equal to time, 
\begin {eqnarray}
  \tfrac 1{(D-1) !} \varepsilon_{\tau i b_1 b_2 \ldots b_{D-1}}
  J_{b_1 b_2 \ldots b_{D-1}}^a &=& \nonumber \\
  \partial_i \partial_\tau u^a - \partial_\tau \partial_i u^a &=& U_i^a, 
\label {Uia}
\end {eqnarray}
isolating the perpendicular direction $i$. Instead of the dislocation
current $J$, we could have used here also the disclination current $T$, but
our interest in this subsection will be in the former.
The comparison of the tensor Eq.(\ref {Uia}) with the glide constraint 
Eq.(\ref {glidestrain}) illustrates that  
the glide condition turns into a constraint on the 
trace $U_i^i = 0$. For rank 2 tensors, the trace is the only invariant 
tensorial component of the $D$-dimensional orthogonal group ($O(D)$), 
corresponding with all rotations and the inversion in 
the $D$-dimensional medium. It follows 
that the constrained  current is the only `singlet' 
(scalar) under the point group 
symmetries of the crystal 
(point group symmetries constitute 
a subgroup of $O(D)$). 
The conjugate degree of freedom must have the same symmetry and 
this can only be compression, the only physical entity being a singlet under
$O(D)$. We have identified the fundamental reason that glide
implies the decoupling of dislocations and compressional stress\cite{ZMN}.

Apart from rotations, the glide constraint is also invariant under 
Galilean space-time translations. Needless to say, the constraint 
does not obey Euclidean Lorentz (space-time) invariance as the time 
direction has a special status both in Eq.(\ref {glidestrain}) and 
Eq.(\ref {glidecurrent}). The origin of this lies, of course, in the 
definition of the crystalline displacement ${\bf u}$ and its role in the 
minimal coupling Eq.(\ref {LEM}). The displacements are defined under 
the assumption that every crystalline site has an equilibrium position 
which implicitly wires in that their world lines extend exclusively in the 
temporal direction. The crystal
defect currents (Eq.(\ref {glideeta})) and the disclination
conservation law of Eq.(\ref {conservationT}) are invariant under
Galilean transformations while they do not
respect invariance under Lorentz boosts.

\section {The Glide constraint, the lattice cut-off and anharmonicity}
\label{GCLCA}

Contrary to our rigorous (`glide only') 
result concerning the linear regime of 
continuum elasticity, in real crystals 
dislocations do climb (albeit at small rates).  
As discussed earlier by two of us\cite{ZMN},
interstitial matter will be exchanged when dislocations collide and this 
process releases climb motions. In what follows, 
we will rederive the glide constraint, 
yet now do so within a fully general framework 
which will enable us to address the
implications of both (higher order) {\em non-linear elasticity} 
and the presence of a lattice cut-off. Higher order corrections
to linear elasticity modify the original glide constraints
of Eq.(\ref{glidecurrent}) giving rise dislocation climb. 
We will leave the detailed analysis of the non-linear problem 
to a later publication \cite {anharmonic}.
In what follows, we outline how the full non-linear theory (including
finite lattice size effects) captures higher order effects
such as climb.

To achieve this aim, we return to the mass continuity equation invoked 
earlier (in unison with Eqs.(\ref{j0.}, \ref{ji.})) yet now,
by examining contributions of higher order derivatives
of the displacement field, we exercise 
far greater care in examining its ramifications. The continuity equation
of Eq.(\ref{gliderho}) implicitly assumes that the 
density and current fields are functionals
of local Eulerian (distorted lattice) coordinates.  
On the other hand, the displacement, 
stress, and other elastic fields are functionals of substantial 
coordinates (i.e. the coordinates defined relative to the undistorted lattice 
coordinates)- the Lagrangian coordinate frame.
As was briefly 
done earlier (Eqs.(\ref{j0.}, \ref{ji.})), we express the local density and 
currents of Eq.(\ref{gliderho}) in terms of volume and velocities
as $\rho = \frac{\rho_{0}}{V}$ and ${\bf j} = \frac {\bf \rho_{0} v}{V}$,
with $\rho_{0}$ the mass of the ideal uniform medium in a unit volume
in the undistorted original medium (i.e. the ideal background mass density). 
Following a distortion, a unit volume element of the original 
medium now occupies a region of volume $V$.
With these relations in tow, 
Eq.(\ref {gliderho}) reads
\begin {eqnarray}
  \partial_\tau V + ({\bf v} \cdot \nabla) V \equiv D_\tau V = 
V (\nabla \cdot {\bf v}). \label {glideEuler}
\end {eqnarray}

This equation can be interpreted as a law governing the change of volume 
of the elastic medium: the change in volume (the derivative on the 
left hand side) is exclusively dictated by the motion of the 
boundaries- the gradient on the right hand side corresponds with a surface 
integral of the velocity field. This is just a reformulation of the same
basic constraint: the conservation of mass (or electrical charge).
Throughout this paper, mass conservation was the primary ingredient 
leading to the glide constraint. To invoke the mass continuity
equation in the form of Eq.(\ref {glideEuler}), let us express the actual
atomic coordinates $R^j$ in terms of the Eulerian coordinates 
(henceforth denoted by $r^i$)
by employing the identity,  

\begin {eqnarray}
  (\frac {\partial R^j}{\partial r^i})_\tau = 
(\delta_{ij} + \sum_{m=1}^\infty \frac {a^{m-1}}{m!} \partial_j^m u^i)^{-1},
\end {eqnarray}
where $a$ is a (lattice) cut-off scale. Simplifying, 
we find that Eq. (\ref{glideEuler}) may be recast as

\begin {eqnarray}
  \partial_\tau V = 
V ~ (\frac {\partial R^j}{\partial r^i})_\tau 
\partial_i \partial_\tau u^j. \label {glideLagr}
\end {eqnarray}

This is an exact expression (entailing corrections to 
all order in gradients of ${\bf u}$) detailing the glide constraint.
Retaining the leading order 
contributions and employing $V = 1 + \partial_i u^i$,
we recover the `familiar' linearized glide constraint
of Eq.(\ref {glidestrain}). The exact glide constraint Eq.(\ref {glideLagr}) 
can also be derived using the coupling to electromagnetism as
generating functional as used in section IV. 

How may we generalize this analysis so that it deals with the 
non-linearities up to all
orders? A key is provided by Eq.(\ref{LEM}). Due 
to changes in the volume of the medium arbitrarily large 
lattice deformations translate into current 
densities $j_{\mu}$.
In addition, we should use substantial- instead of local coordinates
such that the spatial arguments of the EM fields and strains
no longer coincide. A crystalline constituent that was originally at 
position ${\bf R}$ feels the EM field at point ${\bf R} + {\bf u}$.
Hence, the full non-linear generalization of Eq. (\ref{LEM}) is
\begin {eqnarray}
  {\cal L}_{EM} = \frac {(n_e e)}{V ({\bf R})} 
\lbrack \partial_\tau u^i ({\bf R}) A_i ({\bf R} + {\bf u}) 
+ A_\tau ({\bf R} + {\bf u}) \rbrack. 
\label {exactEM}
\end {eqnarray}
Using the procedure of Section IV, the exact glide constraint is 
straightforwardly derived from this expression.

What can we learn from this exact form of the glide
constraint? Let us specialize to the simple 2+1D medium. Assuming a 
vanishing lattice constant $a$, the volume of
an elementary cell is simply given as $V = \det (\delta_{ij} +
\partial_i u^j)$. Inserting this in the exact glide expression 
Eq.(\ref {glideLagr}) we find that it simplifies,
\begin {eqnarray}
  0 = \varepsilon_{ab} J_i^b (\delta_{ia} + \partial_i u^a). 
\label {glide2d1exact}
\end {eqnarray}
The displacement 
derivative $\partial_i u^a$ includes both regular
and singular components. When  $\partial_i u^a$ is 
small compared to unity, the linearized
glide constraint of Eq.(\ref {glidecurrent}) is recovered.

Let us now analyze the  conditions required for 
 a dislocation in climbing motion to satisfy the generalized
constraint of Eq.(\ref {glide2d1exact}). As before, we set the Burgers
vector of a dislocation to ${\bf b} = b {\bf e}_x$ implying
$J_\mu^y = 0$. A climbing dislocation then moves in the $y$ direction
resulting in the dynamical current ($\overline x_i$ is the position
of the dislocation)
\begin {eqnarray}
  J_y^x = b ~ \partial_\tau {\overline y} ~ \delta \lbrack x_i -
  \overline x_i (\tau) \rbrack. \label {Jclimb}
\end {eqnarray}

Assuming that the dislocation does not glide simultaneously, the
current Eq.(\ref {Jclimb}) is the only nonzero dynamical dislocation
current. The exact glide constraint is then satisfied, if and only if
\begin {eqnarray}
  0 = 1 + \partial_y u^y \bigg |_{\overline x_i}. \label {climbstrain}
\end {eqnarray}

Clearly, so long as we keep the displacements small compared than the
lattice size (effectively prohibiting interstitials and
vacancies), the condition of Eq.(\ref {climbstrain}) is impossible to
satisfy and climb motion is strictly forbidden. However, as soon as
particles are allowed to leave their initial positions in the crystal
(interstitial density develops), $\partial_y u^y$ becomes finite 
and climb motion is liberated!

More precisely,
the right hand side of the condition Eq.(\ref {climbstrain}) represents
the derivative of Euler to Lagrangian coordinates $\frac {\partial y}
{\partial Y}$ and if vanishing, it means that two neighboring sites
of the original crystal are, upon deformation, occupying the same
place. This is none other than an interstitial `event'. We can
thus interpret the 2+1-dimensional exact glide 
constraint of Eq.(\ref {glide2d1exact})
as formally proving the old metallurgical maxim that ``climb 
motion is permitted if and only if interstitials/vacancies are present''.
This climb will have the effect that the interstitial is absorbed by
the dislocation.

We reach the conclusion that higher orders in the displacements
(non-linearities, anharmonicities) have 
the effect of restoring the physics of interstitials. Another natural
question to ask regarding the continuum limit
is whether there may be a hidden dependence on the
lattice cut-off $a$. To investigate this issue, let us see what happens
to Eq.(\ref{glideLagr}) when we only keep terms which are
harmonic in the displacements, while imposing no conditions on the number
of derivatives. These have to do with the volume of the unit cell.
 Let us specialize to a simple hypercubic lattice with
sides defined by the vectors  ($i = 1, \ldots, D$),
\begin {eqnarray}
  u^j (R_i + a_i) - u^j (R_i) = \delta_{ij} + \sum_{m=1}^\infty
  \frac {a^{m-1}}{m!} \partial_i^m u^j. \label {uijcubic}
\end {eqnarray}

The corresponding volume is a determinant of the matrix Eq.(\ref {uijcubic})
and in the harmonic approximation only the diagonal elements remain,
\begin {eqnarray}
  V = 1 + 
\sum_{i,j=1}^D \sum_{m=1}^\infty \frac {a^{m-1}}{m!} \partial_i^m u^j.
\end {eqnarray}
Using this expression for the volume, the exact equation of constraint 
(Eq. (\ref {glideLagr})) reads
\begin {eqnarray}
  \varepsilon_{\tau a i_{1} ... i_{D-1}} J^{a}_{i_{1}...i_{D-1}} \nonumber
\\ = - (D-1)! \sum_{i=1}^D \partial_\tau \frac {e^{a \partial_{i}}
  - a \partial_{i} - 1}{a \partial_{i}} \partial_{i} u^{i}. 
\label {glidecubic}
\end {eqnarray}
The remarkable fact is that it is possible to collect all higher order 
derivatives in a simple exponent. 

Eq.(\ref{glidecubic}) amounts to a double check on the non-singular
nature of the continuum limit. It corresponds with a resummation to
all orders and if there would be a non-perturbative surprise it
should show up in this equation. One sees immediately that this 
is not the case: in the long wavelength limit
 ($q \to 0$), the spatial derivatives on the right
hand side vanish recovering the linearized equation of 
constraint (Eq.(\ref {glidecurrent})).

\section {Conclusion}

What has happened in this paper? This pursuit was driven by
a bewilderment with the glide phenomenon. Surely, it is very
easy to explain using the intuitive arguments found in the
elasticity textbook (Fig.1). However, we got puzzled when we
discovered that it plays a rather central and outstanding
role in the field theoretic formulation \cite{ZMN}. It is just not
something that one adds by hand, but instead it is deeply wired
into the structure of the theory, and it is responsible for
the theory to have sensible outcomes like the `prediction'
that non relativistic fluids carry sound.

In this earlier work \cite{ZMN}, it was assumed that glide was at work 
as a prerequisite to find sensible outcomes. In this paper
we nailed down its origin: assuming that mass is conserved
in the non-relativistic elastic medium, dislocations have to
to glide as long as the `interstitial' (non-topological) 
displacements are small compared to the lattice constant. 
In the linearized theory the glide constraint becomes
exact. We learned in section V that this is to an extent
an artifact of the field theory since there is room for
interstitials in the fully non-linear theory. However,
one might want to view the field theory as a Platonic
form, a mathematical ideal, and it is just remarkable
that topology, absence of symmetry (broken Lorentz invariance)
and non-topological conservation laws (mass conservation) 
conspire to produce a kinematic law determined by a
topological invariant.

Our key results in Eqs.(\ref{glidestrain},\ref{glideLagr}) translate mass
conservation into the mathematical language of the topological
defects. These are quite useful because they make possible
to address the workings of the constraint in circumstances
where the visualization `method' of the elasticity books is
of little use. We illustrated this by analyzing dislocation
motions in arbitrary dimensions, demonstrating that this
is about restricted climb (`mass transfer is free inside the
brane'). This is surely of some practical value: we imagine
that it might be useful in the complicated, real life situations
encountered by metallurgists. It will be surely of importance 
in the generalization of the duality by Zaanen, Nussinov and
Mukhin\cite{ZMN} from 2+1D to 3+1D, a task which is still
to be accomplished and not unimportant given the fact that
quantum nematics are more likely be found in a three dimensional
world. A  related issue is that it still remains to be demonstrated
that the Lorentz-invariant version of such a 3+1D quantum nematic
has to do with 4D Einstein gravity\cite{KZ}.    

Why are disclinations not subjected to glide constraints, while
dislocations are? Using our formalism we could address this question
in detail (section IV). In a way, the outcomes are surprising. 
Although we find that indeed the mass conservation does not
impose kinematic constraints on the disclinations, we do
find that they imply a disclination conservation law. Although
at first sight this might appear as obvious, it is actually
quite confusing. Usually topological conservation laws are
rooted in the presence of an order parameter and are thereby 
associated with some generalized rigidity. In this way, dislocations
are about shear rigidity while disclinations are associated 
with curvature rigidity, one of these beautiful facts becoming
clear in the geometrical formulation. Volume and mass are
associated with compressional rigidity. How can it be that
this seemingly non-topological entity volume can still give
rise to conservation laws pertaining to topological currents?

We conclude with a speculation.
As emphasized throughout, the basic physical ingredient leading to
the derived constrained glide dynamics in solids was mass conservation.
In a formal setting, similar albeit 
more restrictive volume conserving diffeomorphisms 
(parameterized by time) may be directly examined via the 
$w_{\infty}$ algebra in two spatial 
dimensions and its extensions, e.g. \cite{witten}. 
We suspect that there might be 
a more fundamental way of casting our relations 
by relying on the intricacies
of such symmetries of space and time.

\section {Acknowledgments}

This work was supported by the Netherlands foundation for fundamental research 
of Matter (FOM). JZ acknowledges the support by the Fulbright
foundation in the form of a senior fellowship. Support
(ZN) from the US DOE under LDRD X1WX at LANL is acknowledged.

\appendix

\section {Dualization and defect densities}

In this appendix, we sketch the dualization procedure for an action 
describing  a three-dimensional elastic
medium. We will then elaborate on the relation between the dual 
fields and the defect densities. The extension of our treatment 
to other dimensions (as well as to imaginary time) is straightforward. 
The dislocation/disclination
densities of Eqs.(\ref{staticdislocation},\ref{staticdisclination}) 
harbor a redundancy which may be removed by
considering a single symmetric tensor encapsulating information
about both dislocations and disclinations- the double 
curl topological defect density $\eta$ of Eq.(\ref{etad}).

Linear elasticity is valid only when the displacements 
are minute as compared to the lattice size. Within this regime, 
the Lagrangian 
density of Eq.(\ref {S0}) is
\begin {eqnarray}
  {\cal L} \lbrack u^a, \partial_i u^a \rbrack = \tfrac 1 2 \partial_i u^a C_{ijab} \partial_j u^b. \label {Llinear}
\end {eqnarray}
This is a functional of displacements $u^a$ and their first 
derivatives $\partial_i u^a$. The elasticity tensor $C$ must 
adhere to symmetries inherited from the underlying lattice. Thus, 
as dictated by the triclinical point symmetry group in 
a three dimensional medium, it may contain up to 21 independent constants. 
[In general dimension $d$, the elastic tensor may have up to
$\tfrac 1 8 d (d + 1) (d^2 + d + 2)$ independent components.]
The displacements ${\bf u}$ have 
generalized momenta conjugate to them-- the stresses given by
\begin {eqnarray}
  \sigma_i^a = \frac {\partial {\cal L}}{\partial (\partial_i u^a)} = C_{ijab}
  \partial_j u^b. \label {sigmaia}
\end {eqnarray}
From the aforementioned symmetry of the elastic tensor $C$, an
Ehrenfest constraint encapsulating a symmetry of the stress tensor 
($\sigma_i^a = \sigma_a^i$) rigorously follows. 
By a Legendre transformation, 
we may obtain the Hamiltonian as functional of displacements and stresses 
\begin {eqnarray}
  {\cal H} \lbrack u^a, \sigma_i^a \rbrack = \sigma_i^a \partial_i u^a - {\cal L}
  \lbrack u^a, \partial_i u^a (u^a, \sigma_i^a) \rbrack. \label {Hlinear}
\end {eqnarray}

As we are after the Lagrangian in its dual 
form, we may first express the Lagrangian
through the displacements and stresses exclusively via
\begin {eqnarray}
  {\cal L}_{dual} = \sigma_i^a \partial_i u^a - {\cal H} \label {Ldual},
\end {eqnarray}
leading to the partition function
\begin {eqnarray}
  Z = \int {\cal D} u^a {\cal D} \sigma_i^a ~ e^{- \int \rmd {\bf x} ~ {\cal L}_{dual}}. \label {Zdual}
\end {eqnarray}

The dualization of the linear 
elastic action proceeds by a complete removal of the displacement
degree of freedom. This is achieved 
by splitting the displacement field into the smooth and singular components 
${\bf u} = {\bf u}_{sm} + {\bf u}_{MV}$. This prescription
generalizes the standard Abelian Higgs duality, e.g. \cite{Zee},
to the elastic arena. 
The singular displacement field ${\bf u}_{MV}$
harbors the defect density while the smooth 
field ${\bf u}_{sm}$ is everywhere regular. Much as in other
dualities \cite{Zee}, the smooth part can be integrated out, leading
a constraint on stresses (momentum conservation)
\begin {eqnarray}
  \partial_i \sigma_i^a = 0. \label {sigmaconserved}
\end {eqnarray}

By virtue of the Ehrenfest constraint, this conservation law 
(Eq.(\ref {sigmaconserved})) is
valid also when the contraction is performed with respect to the second index.
This property enables us to express the stress field as a double curl of the
symmetric two-form $\chi$ (also known as the elastic gauge field 
\cite{Kleinert})
\begin {eqnarray}
  \sigma_i^a = \varepsilon_{ijk} \varepsilon_{abc} \partial_j \partial_b \chi_{k, c}.
\end {eqnarray}
The physical field (stress) is invariant under the gauge transformation
\begin {eqnarray}
  \delta \chi_{k, c} = \partial_k \xi_c + \partial_c \xi_k.
\end {eqnarray}

We are nearly done. The Hamiltonian Eq.(\ref {Hlinear}) may 
now be expressed in terms of the elastic gauge fields $\chi$.
The leftover term with the 
singular displacement $\sigma_i^a \partial_i u_{MV}^a$ leads to
(after integrating by parts)
\begin {eqnarray}
  {\cal L}_{coupl.} = \chi_{k, c} \left \lbrack 2 \Theta_k^c +
  \varepsilon_{abc} \partial_a \alpha_k^b \right \rbrack, \label {Lcoupling}
\end {eqnarray}
with the final identification of the dislocation and disclination
defect densities of Eqs.(\ref{staticdislocation},\ref{staticdisclination}).

As usual (e.g., the famous minimal coupling $(j_{\mu} A_{\mu}$) in
electrodynamics), such densities multiplying a gauge potential
(here $\chi_{k, c}$) serve as source terms. In Eq.(\ref{Lcoupling}),
the charges appearing in the elastic minimal coupling 
are topological. 

The redundancy of the defect densities needed in the coupling term
Eq. (\ref {Lcoupling}) is obvious: from nine independent topological
defect densities (six dislocation density components $\alpha_i^a$ and
six disclination density components $\Theta_i^a$, connected by three
relations), only three linear combinations couple to the physical degrees
of freedom.

To overcome the possible confusion, the coupling term may be 
treated differently
so that the Ehrenfest constraint is utilized at the very start
\begin {eqnarray}
  \sigma_i^a \partial_i u^a = \tfrac 1 2 (\sigma_i^a + \sigma_a^i) \partial_i u^a =
  \sigma_i^a \tfrac 1 2 (\partial_i u^a + \partial_a u^i).
\end {eqnarray}
After the integration by parts, the elastic gauge field $\chi$
becomes minimally coupled to the double curl 
topological defect density of Eq.(\ref {eta3d})
\begin {eqnarray}
  {\cal L}_{coupl.} = \chi_{k, c} \eta_k^c.
\end {eqnarray}

In spite of its usefulness in linear elasticity, the double curl
topological density $\eta$ is insufficient when higher order
derivatives become important and new degrees of freedom are introduced.

\bibliographystyle{apsrev}

\begin{thebibliography}{10}
\bibitem {commentBurgerstau} Frank was the first to pose the question of
a time-like Burgers vector. A natural example is provided by Earth where every point but poles may define a local time (based on the position of Sun). A traveler who makes a trip around the world and notices only smooth changes in time will experience 24h time displacement. Therefore a need for the International Date line that acts as a Volterra cut connecting two singularities (poles).
\end{thebibliography}

\end {document}